\documentclass[useAMS,usenatbib]{mn2e}

\usepackage[breaklinks, colorlinks, citecolor=blue, urlcolor = blue]{hyperref}
\usepackage{graphicx}
\usepackage{amsmath}
\usepackage{amssymb}
\usepackage{ulem}
\def\aw{Alfv\'en}

\title[Non-linear CR diffusion]{Non-linear diffusion of cosmic rays escaping from supernova remnants - II. Hot ionized media}
\author[L. Nava et al.]
{L. Nava$^{1,2,3}$\thanks{E-mail: lara.nava@inaf.it}, 
S. Recchia$^4$,
S. Gabici$^4$,
A. Marcowith$^5$,
L. Brahimi$^5$,
and V. Ptuskin$^6$
\\
$^1$INAF -- Osservatorio Astronomico di Brera, Via Bianchi 46, I-23807 Merate, Italy\\
$^2$INAF -- Osservatorio Astronomico di Trieste, Via Tiepolo 11, I-34131 Trieste, Italy\\
$^3$INFN -- Sezione di Trieste, via Valerio 2, I-34127 Trieste, Italy\\
$^4$APC, AstroParticule et Cosmologie, Universit\'e Paris Diderot, CNRS/IN2P3, CEA/Irfu, Observatoire de Paris, Sorbonne Paris Cit\'e,\\ 10, rue Alice Domon et L\'eonie Duquet, F-75205 Paris Cedex 13, France\\
$^{5}$Laboratoire Univers et particules de Montpellier, Universit\'e Montpellier/CNRS, F-34095 Montpellier, France\\
$^{6}$Pushkov Institute of Terrestrial Magnetism, Ionosphere and Radiowave Propagation, 108840, Troitsk, Moscow, Russia\\
}

\begin{document}
\voffset -1truecm 
\date{}

\pagerange{\pageref{firstpage}--\pageref{lastpage}} \pubyear{}

\maketitle

\label{firstpage}

\begin{abstract}
We study the problem of the escape and transport of Cosmic-Rays (CR) from a source embedded in a fully ionised, hot phase of the interstellar medium (HIM). In particular, we model the CR escape and their propagation in the source vicinity taking into account excitation of Alfv\'enic turbulence by CR streaming and mechanisms damping the self-excited turbulence itself. Our estimates of escape radii and times result in large values (100\,pc, 
$2\times10^5$\,yr) for particle energies $\lesssim20$\,GeV and smaller values for particles with increasing energies (35\,pc and 14\,kyr at 1\,TeV). These escape times and radii, when used as initial conditions for the CR propagation outside the source, result in relevant suppression of the diffusion coefficient (by a factor 5-10) on time-scales comparable with their (energy dependent) escape time-scale. The damping mechanisms are fast enough that even on shorter time scales, the Alfv\'enic turbulence is efficiently damped, and the ratio between random and ordered component of the magnetic field is $\delta B/B_0\ll 1$, justifying the use of quasi-linear theory. In spite of the suppressed diffusion coefficient, and then the increased residence time in the vicinity ($\leq200$\,pc) of their source, the grammage accumulated by CRs after their escape is found to be negligible (at all energies) as compared to the one accumulated while diffusing in the whole Galaxy, due to the low density of the HIM.
\end{abstract}


\section{Introduction}\label{sec:intro}

Galactic cosmic rays (CRs) are believed to be accelerated at supernova remnants (SNRs) via diffusive shock acceleration \citep[see e.g.][for recent reviews]{luke17,eveline}.
In order to test this hypothesis and to reach a full comprehension of the origin of CRs, three crucial aspects of the problem have to be combined: {\it i)} the acceleration of particles, {\it ii)} their escape from the accelerator site, and {\it iii)} their propagation in the interstellar medium (ISM), which regulates their escape from the Galaxy. This paper deals with point {\it ii} in the above list, and describes the escape of CRs from SNRs, and the propagation of such runaway particles in the immediate vicinity of the accelerator site (before they mix with the ubiquitous interstellar CR sea).
We focus here (paper\,II) onto the {\it fully ionized} phase of the ISM, while the case of a partially ionized ISM can be found in a companion paper \citet{nava16} (hereafter paper\,I), itself complemented by an upcoming paper treating warm/cold partially ISM phases (Brahimi et al, paper\,III, in preparation).

The escape of CRs and their confinement in the vicinity of their sources is a non-linear process. This is because CRs excite the magnetic turbulence that in turn scatters and confines them \citep{wentzel}. Both analytical \citep{skilling70,plesser,malkov} and numerical (see paper\,I, \citealt{dangelo}) approaches to the problem can be found in the literature. Here we closely follow the numerical approach described in paper\,I.

The importance of these studies is connected to (at least) two issues. First, the confinement time of CRs in the vicinity of their sources determines the probability to detect gamma rays resulting from hadronic interactions between the runaway CRs and the ambient ISM. The gamma-ray emission is particularly intense if a massive molecular cloud happens to be located in the vicinity of the CR source. The measurements of the gamma-ray intensity and spectrum of clouds located next to SNRs constitute an invaluable tool to test our ideas about CR origin and transport \citep[e.g.][]{atoyan,gabici09,nava13}.
Second, if the self-confinement of CRs in the vicinity of their accelerator sites is effective enough, a significant fraction of the grammage can be accumulated by CRs during their stay in a relatively small region surrounding SNRs \citep[e.g.][]{dangelo}. This fact would radically change the standard picture of CR propagation in the Galaxy, where the grammage is accumulated during the entire time spent by CRs throughout the Galactic disk.

We anticipate here the three main conclusions of this work. First of all, the problem of the confinement of CRs around SNRs must be solved together with the problem of escape of CRs from the accelerator site (because the solution of the latter problem determines the initial setup to be used to solve the former). Second, the confinement time of CRs in a $\sim$100-200\,pc hot fully ionized region surrounding the parent SNR is found to be negligible with respect to the total confinement time of CRs in the Galactic disk.
Third, the effectiveness of damping mechanisms is significant enough to keep the level of magnetic fluctuations at a level $\delta B/B_0 \ll 1$, where $B_0$ and $\delta B$ are the ordered and random component of the magnetic field, respectively.

The paper is structured as follows: in Sec.~\ref{sec:method} we describe the physics and the setup of the problem and in Sec.~\ref{sec:escape} we estimate the escape time of CRs from SNRs. The results from these sections are then used to describe the time evolution of a cloud of runaway CRs and to compute the spectra of CR particles and Alfv\'enic turbulence in the vicinity of the parent SNR (Sec.~\ref{sec:Tim} and \ref{sec:spectra}, respectively). Finally, in Sec.~\ref{sec:residence_time} we compute the grammage accumulated by CRs in the vicinity of their sources and we conclude in Sec.~\ref{sec:conclusions}.

\section{The method}\label{sec:method}
\subsubsection*{Geometry of the system}
We consider the case of 
CRs that escape from the source and are injected in a region embedded in a magnetic field characterised by a large scale ordered component of strength $B_0$ and a random (Alfv\'enic) component of strength $\delta B$.
The latter is described in terms of fluctuations of amplitude $\delta B$ such that:
\begin{equation}
\frac{\delta B^2}{8 \pi} = \frac{B_0^2}{8 \pi} ~\int I(k)\, {\rm d}\ln k ~ ,
\end{equation}
with $I(k)\equiv\delta B(k)/B_0\ll 1$.
In this limit, a flux tube approximation can be adopted to describe the geometry of the system up to distances comparable to the large scale field coherence length $L_{\rm c}$ 
(for a discussion, see \citealt{plesser}).
We limit our analysis to these length-scale and adopt a one-dimensional description of CR transport along $B_0$.

\subsubsection*{Particle diffusion}
The particle diffusion coefficient is determined by the level of streaming instability driven by the particles themselves and by the relevance of damping mechanisms that reduce the amplitude of the turbulence.
According to the resonance condition for the interaction between particles and waves, slab modes with wavelength $k$ interact with particles with Larmor radius 
$r_{\rm L}  \sim 1/k$, 
with $r_{\rm L}=\frac{\gamma m v c}{ZeB_0}$ for a particle of mass $m$, velocity $v$, charge $Ze$ gyrating in a magnetic field of strength $B_0$.
Since particles with energies $>1$\,GeV are considered in this work, we approximate the particle velocity with the speed of light ($v\sim c$).
In the context of quasi-linear theory (justified as long as $I(k)\ll1$) the diffusion coefficient is related to $I(k)$ by the equation:
\begin{equation}
D(E) = \frac{4 ~ c ~ r_{\rm L}(E)}{3 \pi~ I(k)} = \frac{D_{\rm B}(E)}{I(k)} ~ ,
\end{equation}
where $D_{\rm B}(E) = 4cr_{\rm L}/3\pi$ is the Bohm diffusion coefficient.

\subsubsection*{Equations for CR transport and waves evolution}
The equation describing CR transport is:
\begin{equation}
\frac{\partial P_{\rm CR}}{\partial t} + V_{\rm A} \frac{\partial P_{\rm CR}}{\partial z} = \frac{\partial}{\partial z} \left( \frac{D_{\rm B}}{I} \frac{\partial P_{\rm CR}}{\partial z} \right)\,,
\label{eq:CRs}
\end{equation}
where $V_{\rm A}=B_0/\sqrt{4\upi \rho}$ is the \aw\ velocity in an ambient medium of mass density $\rho$, and $P_{\rm CR}$ is the CR partial pressure normalised to the energy density of the ordered component of the magnetic field:
\begin{equation}
\label{eq:partial_pressure}
P_{\rm CR} = \frac{4\pi}{3}v\,p^4\,f(p)\frac{1}{B_0^2/8\pi}\,.
\end{equation}
The coordinate $z$ is taken along the direction of the ordered magnetic field $B_0$, and $z=0$ refers to the center of the CR source.

This equation must be coupled to the equation that describes the evolution of the Alfv\'enic turbulence  (\citealt{skilling1975,mckenzie,bell,malkov,nava16}):
\begin{equation}
\frac{\partial I}{\partial t} + V_{\rm A} \frac{\partial I}{\partial z} = 2(\Gamma_{\rm CR} - \Gamma_{\rm d}) I + Q\,.
\label{eq:waves}
\end{equation}
The term $\Gamma_{\rm CR}$  describes the rate at which waves grow because of CR streaming instability:
\begin{equation}
\label{eq:growth}
2\Gamma_{\rm CR} I = - V_{\rm A} \frac{\partial P_{\rm CR}}{\partial z} ~ .
\end{equation}
The term $\Gamma_{\rm d}$ describes the rate at which mechanisms for wave damping operate. 
We consider two (linear and non-linear) damping mechanisms relevant for the considered phase of the ISM (hot and fully ionized). 
Their nature and explicit expressions will be presented in section~\ref{sec:damping}.

The last term in Eq.~\ref{eq:waves} accounts for an injection of turbulence from an external source (i.e. other than CR streaming) and is set equal to $Q = 2 \Gamma_{\rm d} I_0$.
This choice assures that when the streaming instability is not relevant, the background turbulence settles to a constant level $I = I_0 = D_{\rm B}/D_0$, where $D_0$ is taken to be equal to the average Galactic diffusion coefficient.

In both equations, the time derivative is computed along the characteristic of excited waves:
\begin{equation}
\frac{\rm d}{{\rm} d t} = \frac{\partial}{\partial t} + V_{\rm A} \frac{\partial}{\partial z} ~ .
\end{equation}
The advective terms $V_{\rm A} \partial P_{\rm CR}/\partial z$ and $V_{\rm A} \partial I/\partial z$ have been neglected in the calculations since, as we checked a posteriori, they introduce minor modifications to the solution, according to the  initial conditions of the ambient medium detailed below.

The equations are solved numerically, using a finite difference explicit method.
As boundary conditions we impose that the spatial distribution of CRs is symmetric with respect to $z = 0$, while at $z = \infty$ we set $P_{\rm CR} = 0$ and $I = I_0$.
In fact, the one dimensional treatment proposed here is valid only as far as propagation distances smaller than the field coherence length $L_{\rm c}$ are considered. When the displacement of particles away from their source significantly exceeds $L_{\rm c}$, the transport mechanism switches from one-dimensional to three-dimensional diffusion. As a consequence, the CR density drops quickly after the transition to  
3D diffusion, a behavior similar to that expected in the case of a free-escape boundary for CRs located at a distance of $\sim L_{\rm c}$.
The coherence length of the field is constrained observationally and is of the order of $L_{\rm c} \approx 100$\,pc, though its exact value is quite uncertain.
For this reason, in the following we keep the boundary conditions at $z = \infty$, and we check a posteriori that this assumption is not significantly affecting the results.

\subsubsection*{Initial conditions and properties of the ISM}
The initial conditions for the CR pressure is set as follows:
\begin{eqnarray}
\label{eq:init_condP}
P_{\rm CR} &=& P_{\rm CR}^0 ~~~~~ z < R_{\rm esc}(E) \\
	&=&  0 ~~~~~~~~~ z > R_{\rm esc}(E) ~ ,
\end{eqnarray}
where $R_{\rm esc}(E)$ represents the spatial scale of the region filled by CRs at the time of their escape from the source.
$P_{\rm CR}^0$ is also a function of energy, and is estimated by imposing that the total energy released in CRs is $W_{\rm CR} \sim 10^{50}$\,erg and assuming a given shape for the total CR spectrum (the adopted parameters are summarised below).

As initial condition for the wave turbulence, we impose $I = I_0$ everywhere.
In fact, as suggested by \citet{malkov}, a larger value of $I \gg I_0$ could be chosen as initial condition for $z < R_{\rm esc}$ (to mimic Bohm diffusion inside the accelerator). However, we found that the exact initial value of $I$ inside the source has negligible effect on the solution.\\

All calculations presented in this work have been performed using the following values for the model parameters:
\begin{itemize}
\item a hot, fully ionized medium (HIM) with:
\begin{itemize}
\item total number density $n=0.01$\,cm$^{-3}$ 
\item temperature $T = 10^6$\,K 
\item magnetic field strength $B_0=2\,\mu$G
\end{itemize}
\item escaping CRs with:
\begin{itemize}
\item total energy $W_{\rm CR}=10^{50}$\,erg
\item power-law spectrum from 1\,GeV to 5\,PeV
\item spectral index $g=-2.2$ ($dN/dE\propto E^{g}$)
\end{itemize}
\item{an ambient CR diffusion coefficient equal to $D_{\rm 0}=10^{28} (\frac{E}{\rm10\,GeV}) ^{0.5}$cm$^2$s$^{-1}$}
\end{itemize}
With these values, the Alfv\'en speed is $V_{\rm A}\sim40$\,km\,s$^{-1}$.
\subsection{Damping of Alfv\'en waves}\label{sec:damping}
We introduce here the expressions of the two main Alfv\'en wave damping mechanisms in the HIM: the damping due to the interaction with background turbulence and the non-linear Landau damping.

\subsubsection{Damping by turbulence injected at large scales}

CR self-generated Alfv\'en waves can be damped by the interaction with background turbulence, supposedly injected at large scales by different sources of free energy \citep[e.g.][]{mclow04}. \citet{fg} account for the anisotropy of the turbulent cascade and derive the following damping rate for waves in resonance with particle of Larmor radius $r_{\rm L}$
\begin{equation}
\label{fg04}
\Gamma_{\rm d}^{\rm FG}=  \left(\frac{\epsilon}{ r_{\rm L} V_{\rm A}}\right)^{1/2} \ ,
\end{equation}
where $\epsilon = V_{\rm T}^3/L_{\rm inj}$ represents the energy cascade rate per unit mass. $L_{\rm inj}$ is the turbulence injection scale, fixed at 100\,pc for the HIM phase \citep{yan04}. $V_{\rm T}$ is the rms turbulent velocity at the injection scale. We consider the turbulence at the injection scale to be trans-Alfv\'enic and hence select $V_{\rm T} = V_{\rm A}$. This is likely the case if the main source of turbulence is due to dying supernova remnants with a forward shock becoming trans-sonic and trans-Alfv\'enic. After substituting the expression for $\epsilon$ into Eq.~\ref{fg04} we get:
\begin{equation}
\label{eq:farmer}
\Gamma_{\rm d}^{\rm FG}= \frac{V_{\rm A}}{(r_{\rm L} L_{\rm inj})^{1/2}} \ .
\end{equation}

\subsubsection{Non-linear Landau damping}
Nonlinear Landau (NLL) damping is produced by the interaction of background thermal ions with the beat of two interfering Alfv\'en waves. As it is the case for a single wave in the case of linear Landau damping, ions with speeds slightly slower (faster) than the speed of this beat take (give) energy from the waves. As for a thermal distribution more ions have lower speeds than the beat, the net effect is a damping. We use the expression provided by \cite{weiner13} (see references therein)
\begin{equation}
\Gamma_{\rm d}^{\rm NLL} =~ \frac{1}{2} \sqrt{\frac{\pi}{2}}v_{\rm i}\,k\,I(k)
\end{equation}
where $v_{\rm i}$ is the background ion thermal speed. For a background medium with a temperature T it becomes:
\begin{equation}
\Gamma_{\rm d}^{\rm NLL} = \frac{1}{2}\sqrt{\frac{\pi}{2}\frac{k_{\rm Bol}T}{m_{\rm p}}}\frac{1}{r_{\rm L}}I(k) \ ,
\end{equation}
where $k_{\rm Bol}$ is the Boltzmann constant and $m_{\rm p}$ is the proton mass.

\section{Escape time}\label{sec:escape}
\begin{figure}
\includegraphics[scale=0.45]{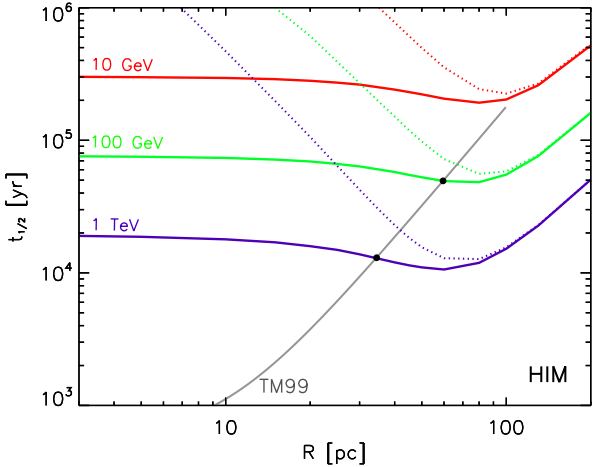}
\caption{Half-time of the CR cloud (see the text, section~\ref{sec:escape}) as a function of its initial radius $R$. Red, green, and purple lines refer to particle energies of 10, 100, and 1000\,GeV, respectively. The grey solid line represents the relationship between SNR radius and age according to \citet{truelove}.}
\label{fig:t1/2}
\end{figure}

Following paper\,I, we define the half-time $t_{1/2}$ of a CR cloud
(see also \citealt{malkov}) and use it to estimate the escape time of CRs of different energies from a SNR. 
The half-time is defined as the time after which half of the CRs initially confined within a region of size $R$ have left the region.
The half-time is then a function of the initial radius $R$ and of the particle energy $E$.
The procedure adopted to estimate $t_{1/2}$ is the following. 
For each energy $E$ we assume that CRs are initially distributed homogeneously inside a region of radius $R$. We solve equations~\ref{eq:CRs} and \ref{eq:waves} considering this configuration as initial condition and find the time after which half of the particles are still trapped inside the initial region size $R$, while the other half has escaped. We repeat the procedure for different values of the initial radius $R$.
The results are shown in Figure~\ref{fig:t1/2} (solid curves), where $t_{1/2}$ as a function of the initial radius $R$ is shown for particle energies of 10\,GeV (red line), 100\,GeV (green line), and 1\,TeV (purple line).

To understand the individual role of the two different damping mechanisms, we also show the results in case only $\Gamma_{\rm d}^{\rm FG}$ is included (dotted curves).
At small radii, the large CR gradient entails a large amplification of \aw\ waves and as a consequence $\Gamma_{\rm d}^{\rm NLL}$  (that is proportional to $I$) dominates over $\Gamma_{\rm d}^{\rm FG}$.
At intermediate radii, the two damping rates are comparable.
Larger radii correspond to a reduced efficiency in the amplification of \aw\ waves and the test particle regime is recovered (i.e., $t_{1/2} \approx R^2/D_0$).

The half-time of the CR cloud gives a rough estimate of the escape time of CRs from a region of initial size $R$. 
The size of a SNR ($R_{\rm SNR}$) and its evolution with time can be estimated, for an SNR expanding in a homogeneous medium with density $n$ as \citep{truelove,ptuskinzirakashvili2005}:
\begin{equation}
\label{eq:Rs}
R_{\rm SNR} = 5.0~ \left( \frac{E_{\rm SNR,51}}{n} \right)^{1/5} \left[ 1 - \frac{0.09 M_{ej,\odot}^{5/6}}{E_{\rm SNR,51}^{1/2} n^{1/3} t_{\rm kyr}} \right]^{2/5} t_{\rm kyr}^{2/5} ~ {\rm pc}
\end{equation}
where $E_{\rm SNR,51}$ is the supernova explosion energy in units of $10^{51}$~erg, $n$ the total density of the ambient ISM in cm$^{-3}$, $M_{\rm ej,\odot}$ the mass of the supernova ejecta in solar masses, and $t = 10^3 t_{\rm kyr}$\,yr is the SNR age.

The equation above is valid for times longer than $\approx 0.4 ~E_{\rm SNR,51}^{-1/2} M_{\rm ej,\odot}^{5/6} n_{-2}^{-1/3}$\,kyr, while for earlier times an appropriate expression for the free expansion phase of the SNR evolution must be used \citep{chevalier1982}. Also, the validity of equation~\ref{eq:Rs} is limited to times shorter than $\approx 2 \times 10^5 E_{\rm SNR,51}^{3/14}/n_{-2}^{4/7}$ yr, which marks the formation of a thin and dense radiative shell \citep{cioffi}.

Equation~\ref{eq:Rs} is plotted (within the limits of its validity) in Fig.~\ref{fig:t1/2} as a grey line labelled TM99 for $E_{\rm SNR,51} = M_{\rm ej,\odot} = 1$ and $n = 0.01$\,cm$^{-3}$.
For each energy $E$, we define the CR escape time $t_{\rm esc}(E)$ and radius $R_{\rm esc}(E)$ from the intersection between the $t_{1/2}$ curves and the TM99 curve.
Within this framework, we can reproduce the qualitative result that higher energy particles escape SNRs earlier than lower energy ones \cite[see e.g.][and references therein]{gabiciescape}.
More precisely, 1\,TeV particles are found to escape when $R_{\rm SNR}\approx 35$\,pc, 100\,GeV particles escape at $R_{\rm SNR}\approx 60$\,pc (corresponding approximately to SNR ages of $1.4\times10^4$\,yr and $5\times10^4$\,yr, respectively).
For low energy particles ($E\lesssim20$\,GeV), the definition of the escape time is more problematic. 
Here we assume that they are released at once, when the Mach number is of the order of a few (i.e., at $\approx2\times10^5$\,yr, when $R_{\rm SNR}\approx100$\,pc), which also roughly corresponds to the onset of the radiative phase. Meanwhile, due to the high ambient sound speed in the HIM, the SNR becomes trans-sonic at the end of the Sedov phase \citep{parizot}.
Moreover, in the radiative phase the mechanism for particle escape is not well understood and other scenarios are possible. For instance, CRs may be released progressively, like high energy particles. In all these cases the estimate of the escape time of GeV particles does not change dramatically.

We can compare these findings with the results from \cite{nava16}, where we considered the partially ionized phases of the ISM. For a supernova exploding in the warm ionized medium (of total density $n = 0.35$ cm$^{-3}$ and a ionization fraction equal to 0.9), we found that CRs of energy 1\,TeV, 100\,GeV, and 10\,GeV leave the SNR when its radius is $\approx$ 11, 16, and 30\,pc, respectively. This corresponds to SNR ages of $\approx$ 4.7, 12, and 51\,kyr. 
For the warm neutral medium (total density $n = 0.35$ cm$^{-3}$ and ionization fraction equal to 0.02) and for the same particle energies the inferred escape radii are $\approx$ 8, 15, and 28\,pc, corresponding to SNR ages of $\approx$ 2.1, 9.5, and 44\,kyr, respectively.
The escape times (radii) are shorter (smaller) when the partially ionized phases of the ISM are considered. This is due, on one hand, to the larger gas densities which hinders the expansion of the SNR and, on the other, to the effectiveness of ion-neutral damping of Alfv\'en waves which balances the effects of streaming instability and allows CRs to escape more easily \citep{zweibel82}.

As a final remark, we note that the estimates of the cloud half-times $t_{1/2}$ as a function of the initial radius (solid curves in Fig.~\ref{fig:t1/2}) are affected by a different choice of $D_0$ only at large radii, where the diffusion is well approximated by a test particle regime and $t_{1/2}\propto R^2/D_0$. At smaller radii, where the diffusion is dominated by the self-excited turbulence, $t_{1/2}$ is not affected by the value chosen for $D_0$. 
Since the escape times and radii (black filled circles in Fig.~\ref{fig:t1/2}) fall in the regime marking the transition between the self-excited turbulence dominated regime and the background turbulence dominated regime, they are only moderately affected by the choice of $D_0$. As an example, a $D_0$ ten times larger results in an escape time 20\% shorter as compared to the results obtained with the average Galactic value, and then in smaller escape radius (8-10\% smaller).

\section{Time evolution of a cloud of cosmic rays released by a supernova remnant}\label{sec:Tim}

\begin{figure*}
{
\includegraphics[scale=0.43]{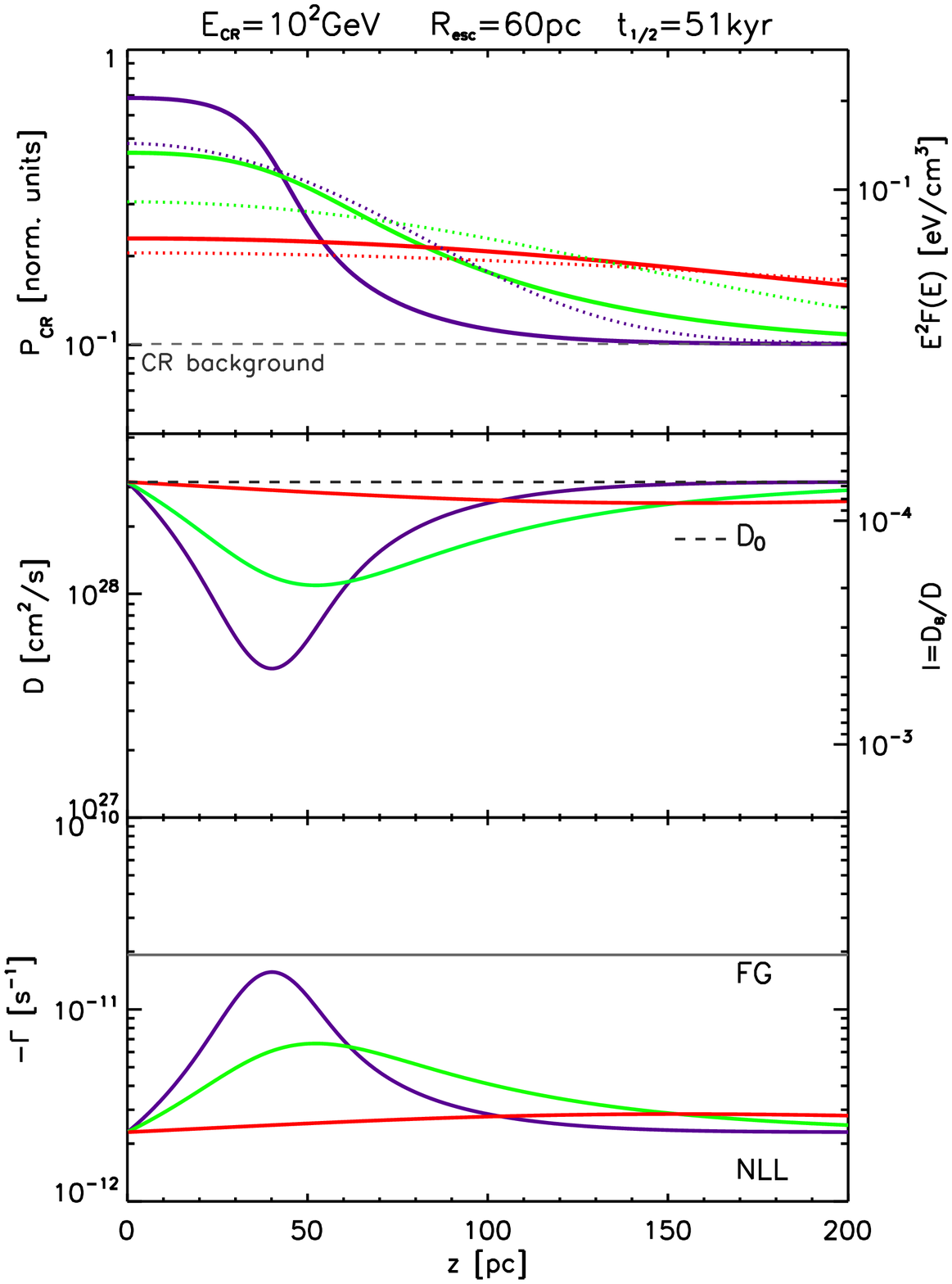}
\includegraphics[scale=0.43]{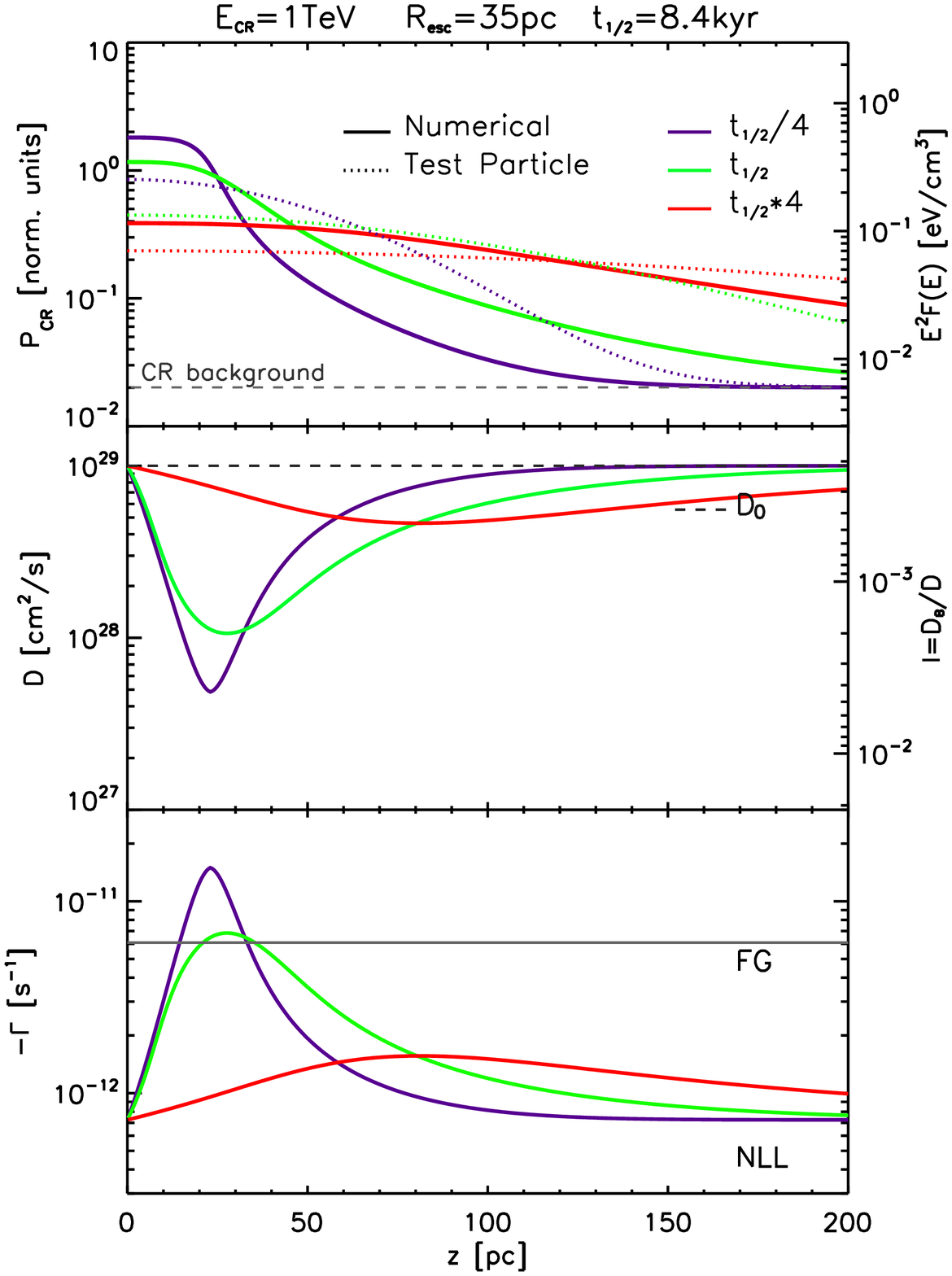}
}
\caption{Evolution in time of the spatial distribution of a CR cloud with initial radius $R_{\rm esc}$ in the HIM phase. $z$ is the distance from the center of the SNR.
CR energies of 100\,GeV (left) and 1\,TeV are considered (right). 
For each energy, the top panel shows the CR partial pressure $P_{\rm CR}$ (see definition in equation~\ref{eq:partial_pressure}), also in terms of CR energy density (see right-hand axis).
The middle panel shows the diffusion coefficient $D$ (left-hand axis) and $I={D_B/D}$ on the right-hand axis.
The bottom panel shows the rate of the damping mechanisms (see section~\ref{sec:damping}). In each panel the solid curves show the solution to equations~\ref{eq:CRs} and \ref{eq:waves} at three different times, as labeled in the upper left panel. Dotted lines show the test particle solution.}
\label{fig:clouds}
\end{figure*}

For each energy, equations~\ref{eq:CRs} and \ref{eq:waves} can now be solved adopting for $R_{\rm esc}$ the escape radius derived in the previous section.
The solutions for at three different times for two different energies ($E_{\rm CR}=100$\,GeV and $E_{\rm CR}=1$\,TeV) are show in Fig.~\ref{fig:clouds}.
The particle energy, escape time and escape radius are reported in the title of each figure.
The upper panel shows the CR pressure as a function of the distance from the SNR centre.
The $y-$axis on the left refers to normalised (to the magnetic field energy density) units, and the $y-$axis on the right is the particle energy density in physical units.
The panel in the middle shows the CR diffusion coefficient $D$, also in terms of ratio $I=D_{\rm B}/D$ (right $y$-axis).
The bottom panel shows the rate of the damping mechanisms.
In all panels, the three different curves refer to the solution of equations~\ref{eq:CRs} and \ref{eq:waves}  at three different times (estimated from the escape time).
Purple curves refer to early times, corresponding to $t_{1/2}/4$, when most of the particles are still in the initial region.
Green and red solid lines show the solution at times equal to $t_{1/2}$, and $4\times t_{1/2}$, respectively.
The dotted curves refer to the test-particle solution of the problem: these have been obtained considering the same initial conditions and geometry as for the numerical case, but the solution is evolved considering a constant diffusion coefficient, fixed to its average Galactic value $D_0$.
Finally, the dashed horizontal grey lines in the upper and middle panels represent the level of the CR background in the Galaxy and the average diffusion coefficient $D_0$, respectively.
The Galactic CR background has been estimated from \cite{aguilar15}.
Note that the solution is shown up to distances (from the center of the remnant) of 200\,pc. 
Solutions at these large distances must be taken with care, because we are not considering perpendicular transport and then they are valid only for coherence lengths $L_c \lesssim 200$\,pc.
\\
\noindent
The results can be summarised as follows:
\begin{itemize}
\item at early times $t\lesssim t_{1/2}$, the distribution derived considering the role of streaming instability clearly differs from the test-particle solution. The self-confinement produces an increased CR pressure at small distances and a depletion of particles at larger distances, as compared to the test particle solution.
The solution approaches the test particle solution at times significantly longer than the cloud half-time (red curves).
 This implies that $t_{1/2}$ represents an order of magnitude estimate of the time interval during which waves can grow significantly above the background level (middle panels) in a region surrounding the initial CR cloud. This is an energy dependent effect, since $t_{1/2}$ is a decreasing function of particle energy. Thus, for CR energies of the order of 1\,TeV or above, relevant for ground-based $\gamma$-ray observations, the growth of waves operates for about ten thousands years;
 
\item large excesses of CRs above the galactic background can be maintained for times much longer than $t_{1/2}$. This is a well known result from the test-particle theory \citep[e.g.][]{atoyan,gabici09} which can be easily verified after comparing the values of the CR partial pressure (see scale on the right $y$-axis in Fig.~\ref{fig:clouds}) with the total energy density of CRs in the galactic disc, which is indicated by a dashed black line in the top panel of Fig.~\ref{fig:clouds};

\item the strong gradient of CRs close to the cloud edges and the consequent growth of \aw\ waves result in a sizeable suppression of the diffusion coefficient with respect to its Galactic value $D_0$. 
The suppression remains significant in a region of several tens of parsecs surrounding the SNR. The ratio $D_{\rm B}/D$ reaches at most a value $<10^{-2}$ (see right-hand $y$-axis in the middle panels), justifying the initial assumption of quasi-linear theory;

\item since CRs with higher energies are assumed to escape at earlier times, when the SNR radius is smaller, their diffusion coefficient at the time of their escape is suppressed by a larger factor, as compared to low energy particles. This makes the non-linear Landau (NLL) damping particularly relevant for high energy particles, and less relevant for lower energies (as evident also from the comparison between solid and dotted curves in Fig.~\ref{fig:t1/2}), but still comparable to $\Gamma_{\rm d}^{\rm FG}$ at times comparable with the particle escape time;

\item the results presented in Fig.~\ref{fig:clouds} have been derived assuming a CR spectral index $g=-2.2$. The spectral shape of escaping particles is expected to have some impact on the results. As an example, we discuss the case of a steeper value: $g=-2.4$. We find of course that the impact on the results is stronger at large energies and becomes progressively smaller for lower energies. For $E_{\rm CR}=1$\,TeV, the escape time is now $t_{\rm esc}=8.4$\,kyr ($R_{\rm esc}=29$\,pc). With these initial conditions, the spatial distribution of the CR cloud at 1\,TeV and its evolution in time is very similar to the one depicted in Fig.~\ref{fig:clouds} (right-hand panel), but the reference time-scale $t_{1/2}$ is smaller.

\end{itemize}

\begin{figure}
\hskip -0.8truecm
\includegraphics[scale=0.51]{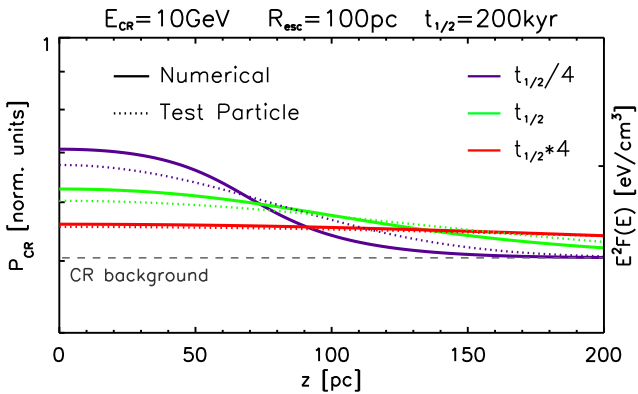}
\caption{Evolution in time of the spatial distribution of a cloud of CRs with energy $E_{\rm CR}=10$\,GeV and initial radius $R_{\rm esc}=100$\,pc in the HIM phase. $z$ is the distance from the center of the SNR.
The solid curves show the CR partial pressure $P_{\rm CR}$ (see definition in equation~\ref{eq:partial_pressure}) at different times, also in terms of CR energy density (see right-hand axis).
Dotted lines show the test particle solution.}
\label{fig:clouds_10GeV}
\end{figure}

For completeness, we also show the evolution of the CR cloud at 10\,GeV (Fig.~\ref{fig:clouds_10GeV}). Given the large radius at which these particle are released in our modeling, they only moderately excite self-generated turbulence, and the spatial distribution and its evolution in time (solid lines) are very similar to those obtained in the test-particle scenario (dotted curves). The diffusion coefficient (not shown) is suppressed only at times smaller than the $t_{1/2}$ time-scale (a factor of 3 at $\sim$\,50\,kyr), and NLL damping plays only a minor role (-$\Gamma^{\rm NLL}_{\rm d}=7\times10^{-12}-2\times10^{-11}$\,s$^{-1}$) as compared to the FG damping (-$\Gamma^{\rm FG}_{\rm d}=6\times10^{-11}$\,s$^{-1}$).

\section{Particle and Turbulence spectra}\label{sec:spectra}
\begin{figure*}
\includegraphics[scale=0.65]{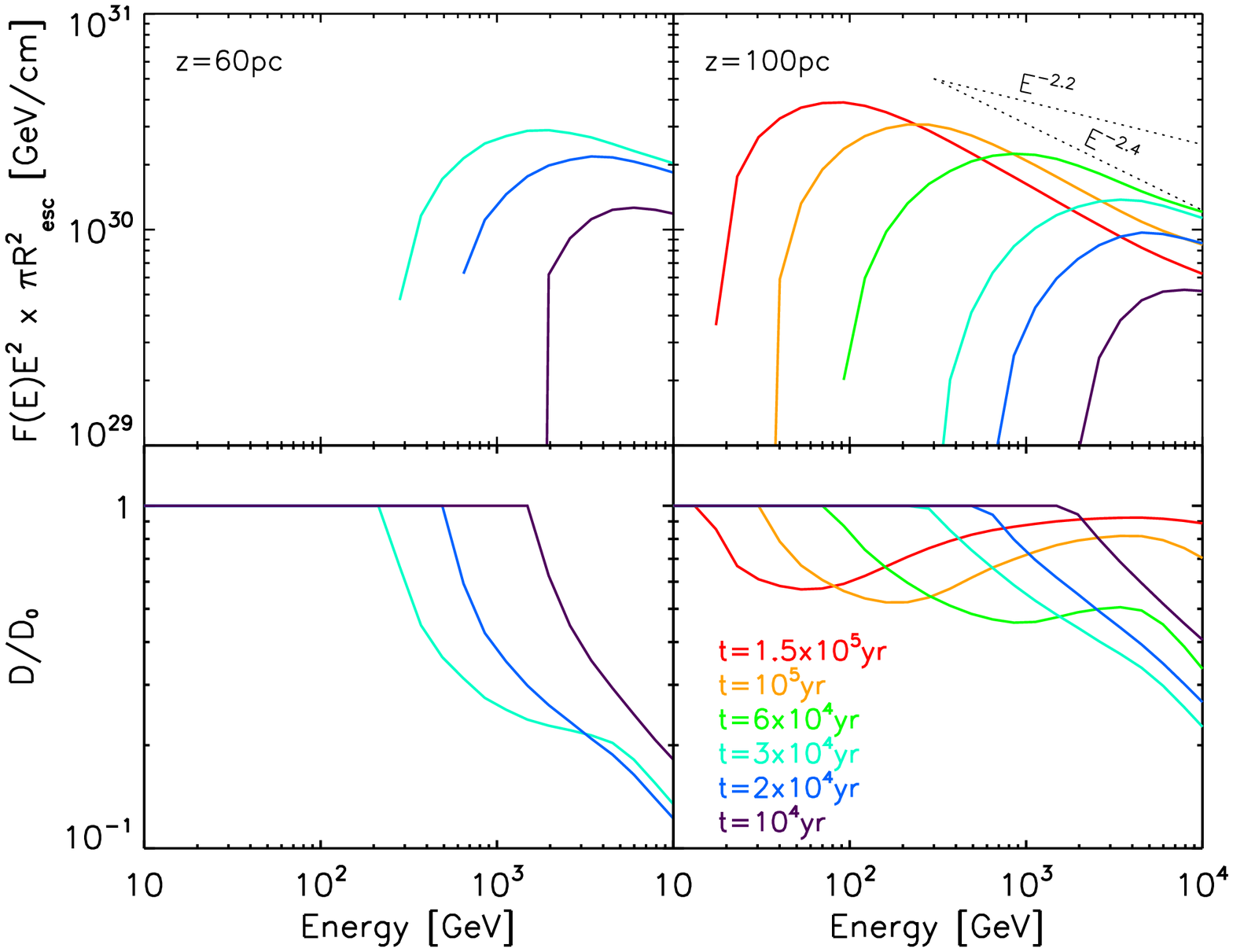}
\caption{Top panels: spectra.  Bottom panels: ratio between the diffusion coefficient and its average Galactic value. Different colours refer to different times since explosion (see the panel on the bottom-right). Left-hand (right-hand) panels refer to a distance of 60\,pc (100\,pc) from the SNR center. The dotted lines in the upper-right panel show power-laws with indices -2.2 and -2.4 (for the CR injection spectrum, a value $g=-2.2$ has been assumed). }
\label{fig:spectra}
\end{figure*}

The spectral shape of CRs released by the source and the amount of Alfv\'enic turbulence produced by CR streaming will strongly depend on the location (i.e., distance from the source) and time.
Fig.~\ref{fig:spectra} shows the spectra (upper panels) and the deviation of the diffusion coefficient from its background value (bottom panels) at two different distances (estimated from the center of the remnant): $z=60$\,pc (left-hand panels) and $z=100$\,pc (right-hand panels). 
In each panel, different curves refer to different times since the SN explosion, from $t=10^4$\,yr to $t=1.5\times 10^5$\,yr (see the legend in the bottom right-hand panel). \\
For a given distance, only solutions at times smaller than the time at which the remnant reaches that distance are shown.

To remove the effect of the simple geometry that we are considering, we have multiplied the spectra by $\pi\,R^2_{\rm esc}$, where the escape radius $R_{\rm esc}(E)$ is a function of the CR energy. All spectra are characterised by a steep rise at low energies, a peak, and then by a decline very close to a power law at particle energies above the peak.

A first result that can be inferred from the inspection of spectra is, as expected, that CRs of lower energy reach a given distance at later times as compared to those of higher energy, that diffuse faster and are released at earlier times.
Concerning the spectral index, in the top-right panel, the two dotted lines show, for comparison, a power-law with index equal to the spectral index of the injected spectrum ($g=-2.2$) and a power-law with a steeper index $-2.4$. Above the peak, all the computed spectra are steeper than the spectrum of released CRs. Again, this is a consequence of the faster diffusion of high energy particles.

The bottom panels in Fig.~\ref{fig:spectra} show, for the same fixed distances $z=60$\,pc (left) and $z=100$\,pc (right), the ratio $D(E,t,z)/D_0$ as a function of energy. Different curves show the evolution in time.
As it can be deduced from the left-hand panel, at a fixed time the derived diffusion coefficient $D$ is equal to $D_0$ at low energies, because low energy CRs have not been released yet or have not reached yet the location where $D$ is estimated. 
At higher energies, the value of $D$ starts to deviate from $D_0$. At later times, the energy above which a deviation is evident decreases.

A somewhat more complex behavior is governing the shape of the curves at $z=100\,$pc (right-hand bottom panel).
Let us focus, as an example, on the curve at $t=10^5\,$yr: the turbulence produced by particles below $\sim$\,30\,GeV has not yet reached a distance of 100\,pc, explaining why $D(E<30\,\rm GeV,10^5\,\rm yr,z=100\,\rm pc) = D_0(E<30\,\rm GeV)$. At higher energies, $D$ starts to deviate from $D_0$. At even larger energies, the turbulence produced by high energy particles has already diffused to larger distances and is being damped by mechanisms that decrease its value so that $D$ tends to its background value. A peculiar behavior is observed at even higher energies ($E\gtrsim4$\,TeV): the level of turbulence increases again, producing another decrease in the $D(E)/D_0$ curve. This behavior is a direct consequence of how the escape radius depends on the particle energy. The function $R_{\rm esc}(E)$ is indeed a decreasing function of $E$, well described by a broken power-law behavior. The break appears around 4\,TeV and at higher energies the escape radius is a stronger function of the CR energy. This means that at energies larger than 4\,TeV the escape radius is small enough to trigger streaming instabilities at a larger rate (remind that the density of escaping CRs is proportional to a dilution factor of $1/\pi R_{\rm esc}$), that reflects in a larger amount of Alfv\'enic turbulence.
This energy-dependent efficiency of the particle self-generated turbulence (with high-energy particles confining themselves more efficiently) results in a larger density of the high-energy particles.
This explains the hardening of the particle spectra, visible in the upper-right panel at TeV energies.

\section{Residence Time and grammage}\label{sec:residence_time}
The amount of time that CRs spend close to their source is a crucial quantity to understand whether the grammage accumulated by particles while diffusing in the source vicinity is a relevant or negligible fraction of the total grammage accumulated while diffusing in the whole Galaxy.
 
Given the low density of the ISM phase considered in this work, we expect this contribution to be negligible in such systems. Due to the relevant suppression of the diffusion coefficient around the source, that will results in a longer residence time, it is however still interesting to perform an estimate of such a contribution.

A formal derivation of the grammage should rely on the solution of the transport equation (Eq.~\ref{eq:CRs}) for CR nuclei, with an additional term describing nuclear fragmentation (see e.g. \citealt{berezinskii90} and \citealt{ptuskin90}). 

\begin{figure}
\includegraphics[scale=0.475]{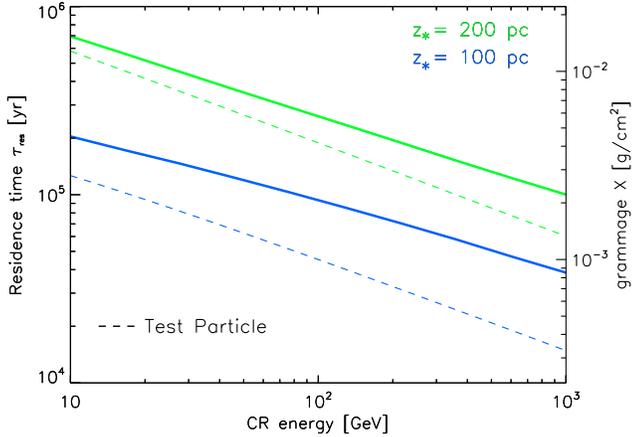}
\caption{Solid curves: residence time $\tau_{\rm res}$ within a region $z_*=100$\,pc (blue) and $z_*=200$\,pc (green) as a function of particle energy. Dashed curves: same quantity, derived in the test-particle approximation.}
\label{fig:residence_time}
\end{figure}

We propose here a simpler, order of magnitude estimate of the grammage.
We first compute at every time $t$ the average diffusion length: 
\begin{equation}
\langle z^2(t)\rangle= \frac{\int_0^\infty z^2 P_{\rm cr}(t,z, E)\,dz}{\int_0^\infty P_{\rm cr}(t,z, E)\,dz}\,.
\end{equation}
This definition is justified because advection is found to be negligible.
Then we define the residence time $\tau_{\rm res}(E, z_*)$, within a region of size $z_*$, as the time such that:

\begin{equation}
\langle z^2 (\tau_{\rm res})\rangle = z_*^2
\label{eq:residence_time}
\end{equation}
Note that the time computed using equation~\ref{eq:residence_time} in fact also includes the time spent inside the source, and can be then considered as an upper limit to the residence time.
Its value for $z_*=100$\,pc and $z_*=200$\,pc and for CR energies between 10\,GeV and 1\,TeV is shown in Fig.~\ref{fig:residence_time} with solid lines. Dashed lines show the results for the test-particle case. The solutions for the test-particle case have been derived assuming the very same initial conditions (i.e. escape radius and time) adopted for the (numerical) solution including streaming instability.

As expected, the largest residence times are obtained for low energy particles. 
The grammage, estimated from $X\approx1.4\,m_{\rm p}\,n\,c\,\tau_{\rm res}$, (where we have considered a medium composed of 10\% of Helium) is reported on the $y$-axis on the right, in the same plot, and is more than two orders of magnitude smaller that the value of the grammage inferred from observations (not surprisingly, given the very low density of the HIM).

An estimate of the amount of grammage accumulated near the CR accelerators when CR self-confinement is taken into account has been recently presented in \cite{dangelo}. For a HIM with density $n=0.01$\,cm$^{-3}$, and for a distance of $z_*=100$\,pc their estimate is larger than what reported in this work by a factor of $\sim6-7$. Likely, this is mostly caused by a difference in the assumed value of CR acceleration efficiency (10\% in this work and 20\% in \citealt{dangelo}), and on the fact that \citet{dangelo} assumed that all CRs are released at the same time when the SNR radius is equal to 20\,pc. This induces an enhancement of the streaming instability at low energies when compared to our results, where the release time and SNR radius for CRs of a given energy are computed as described in Sec.~\ref{sec:escape}, and are in general larger than 20\,pc. In both cases, however, the conclusion is similar: the low density of the considered medium prevents a sizeable contribution from regions close to the CR source.

On the other hand, \citet{dangelo} claim that a significant fraction of the grammage can be accumulated by CRs in the vicinity of their sources, if a denser and partially ionized ISM is considered. This differs from our earlier results presented in paper\,I, mainly due to very different assumption made about the damping mechanism of Alfv\'en waves. Such an important issue will be briefly discussed in the next Section, and described in detail in 
a forthcoming publication, as in this paper we limit ourselves to consider the hot and fully ionized phase of the ISM.

\section{Discussion and conclusions}\label{sec:conclusions}

In this paper, we discussed the problem of the escape of CRs from SNRs. The presence of runaway particles induces an amplification of the magnetic (Alfv\'enic) turbulence in the region surrounding the SNR. The amplification mechanism is CR streaming instability. Given that CRs are confined in the vicinity of SNRs by the turbulence they themselves generate, the problem is highly non-linear. 
Here, we limited ourselves to the case of a SNR exploding in the hot, fully ionized phase of the ISM, while the case of the partially ionized ISM was considered in paper\,I. 

An important conclusion from the work reported in this paper which was also found in paper\,I 
is the fact that the problem of CR escape from SNRs and that of the CR-induced amplification of the magnetic field in the SNR neighbourhood are intimately interconnected and have to be solved together. This is because the determination of the escape time of CRs defines the radius of the SNR at the time of escape ($R_{\rm esc}(E)$), which in turn regulates the effectiveness of the amplification of the magnetic turbulence induced by runaway CRs (through a dilution factor scaling as $\propto 1/R_{\rm esc}(E)^2$ in the growth rate).
Remarkably, for all the cases considered, we found out that CRs escaping from a SNR at a time $t_{\rm esc}(E)$ after the supernova explosion can significantly amplify the turbulent magnetic field in a surrounding region of size of the order of $\lesssim$ 100\,pc for a time which is also roughly of the order of $t_{\rm esc}(E)$.
Since the escape time $t_{\rm esc}(E)$ is always much smaller than the confinement time of CRs within the Galactic disk $\tau_{\rm res}^{d}$ (millions of years for CRs in the GeV energy domain), we can conclude that CRs spend typically a very short fraction of $\tau_{\rm res}^{d}$ in the vicinity of SNRs. It follows that also the grammage accumulated close to the SNR is very small.

In principle, the grammage accumulated by CRs in the vicinity of their sources could be much larger if a denser phase of the ISM is considered. The warm phase of the ISM would satisfy this requirement, being characterised by a typical density in the range $0.1 - 1$ cm$^{-3}$). 
However, as discussed in \citet{plesser} and in paper\,I, the presence of neutrals would introduce an additional and very effective damping mechanism, namely, ion-neutral friction, which would prevent the magnetic turbulence to grow to the level needed to confine CRs long enough.
\citet{dangelo} pointed out that the effect of ion-neutral friction could be overcome in a medium made of almost entirely ionized hydrogen with a 10\% component of almost entirely neutral He, because the charge exchange cross section for H$^+$-He collisions is dramatically reduced with respect to H$^+$-H collisions (which are generally adopted to compute the ion-neutral damping coefficient). Even though the composition of such a medium would closely resemble that of the warm ionized phase of the ISM \citep{ferriere}, the argument presented in \citet{dangelo} might not be completely correct, because ion-neutral friction is not due exclusively to charge-exchange, but also to elastic ion-neutral collisions, which can also contribute to momentum exchange between ions and neutrals \citep[see e.g.][]{daniele}. A more detailed study in this direction will be presented elsewhere.

The results derived in this paper apply to the case of SNRs expanding in a diluted, hot, and completely ionized ISM. This could be the case, for example, of a SNR located at large Galactic latitudes or inside a superbubble. 
However, the value of the magnetic field coherence length adopted here ($L_{\rm c} \approx 100$\,pc) could overestimate its typical value in superbubbles. For instance, the field in superbubbles may be turbulent (see~\citealt{bykov01}). The most important effect of a significantly shorter field coherence length would be to induce an early transition from the one-dimensional to three-dimensional diffusion of CRs.


\section*{Acknowledgements}
LN acknowledges funding from the European Union's Horizon 2020 Research and Innovation programme under the Marie Sk\l odowska-Curie grant agreement n. 664931.
SR acknowledges support from the region \^Ile-de-France under the DIM-ACAV programme.
SG acknowledges support support from Agence Nationale de la Recherche (grant ANR- 17-CE31-0014) and from the Observatory of Paris (Action F\'ed\'eratrice CTA).

\label{lastpage}

\end{document}